\documentclass[aps,prb,twocolumn,reprint,superscriptaddress]{revtex4-2}
\UseRawInputEncoding 
\usepackage{graphicx}
\usepackage{amsmath}
\usepackage{amssymb}
\usepackage{colordvi}
\usepackage{mathrsfs}
\usepackage{bm}
\usepackage{verbatim}
\usepackage{dcolumn}
\usepackage{multirow}
\usepackage{rotating}
\usepackage{booktabs}
\usepackage{epsfig}
\usepackage{subfigure}
\usepackage{makecell}
\usepackage[colorlinks,allcolors=blue]{hyperref}

\begin{document}
\title{Valley-polarized quantum anomalous Hall effect in van der Waals heterostructures based on monolayer jacutingaite family materials}

\author{Xudong Zhu$^\S$}
\affiliation{ICQD, Hefei National Research Center for Physical Sciences at the Microscale, University of Science and Technology of China, Hefei, Anhui 230026, China}
\affiliation{CAS Key Laboratory of Strongly-Coupled Quantum Matter Physics, and Department of Physics, University of Science and Technology of China, Hefei, Anhui 230026, China}
\author{Yuqian Chen$^\S$}
\affiliation{ICQD, Hefei National Research Center for Physical Sciences at the Microscale, University of Science and Technology of China, Hefei, Anhui 230026, China}
\affiliation{CAS Key Laboratory of Strongly-Coupled Quantum Matter Physics, and Department of Physics, University of Science and Technology of China, Hefei, Anhui 230026, China}

\author{Zheng Liu}
\email[Correspondence author:~]{zhengliu@mail.ustc.edu.cn}
\affiliation{ICQD, Hefei National Research Center for Physical Sciences at the Microscale, University of Science and Technology of China, Hefei, Anhui 230026, China}
\affiliation{CAS Key Laboratory of Strongly-Coupled Quantum Matter Physics, and Department of Physics, University of Science and Technology of China, Hefei, Anhui 230026, China}

\author{Yulei Han}
\email[Correspondence author:~]{han@fzu.edu.cn}
\affiliation{Department of Physics, Fuzhou University, Fuzhou, Fujian 350108, China}

\author{Zhenhua Qiao}
\email[Correspondence author:~]{qiao@ustc.edu.cn}
\affiliation{ICQD, Hefei National Research Center for Physical Sciences at the Microscale, University of Science and Technology of China, Hefei, Anhui 230026, China}
\affiliation{CAS Key Laboratory of Strongly-Coupled Quantum Matter Physics, and Department of Physics, University of Science and Technology of China, Hefei, Anhui 230026, China}

\date{\today}

\begin{abstract}
  We numerically study the general valley polarization and anomalous Hall effect in van der Waals~(vdW) heterostructures based on monolayer jacutingaite family materials Pt$_{2}$\textit{AX}$_{3}$ ($ A $ = Hg, Cd, Zn; $ X $ = S, Se, Te). We perform a systematic study on the atomic, electronic, and topological properties of vdW heterostructures composed of monolayer Pt$_{2}$\textit{AX}$_{3}$ and two-dimensional ferromagnetic insulators. We show that four kinds of vdW heterostructures exhibit valley-polarized quantum anomalous Hall phase, i.e., Pt$ _{2}$HgS$_{3}$/NiBr$_{2}$, Pt$_{2}$HgSe$_{3}$/CoBr$_{2}$, Pt$_{2}$HgSe$_{3}$/NiBr$_{2}$, and Pt$_{2}$ZnS$_{3}$/CoBr$_{2}$, with a maximum valley splitting of 134.2~meV in Pt$_{2}$HgSe$_{3}$/NiBr$_{2}$ and sizable global band gap of 58.8~meV in Pt$ _{2} $HgS$ _{3} $/NiBr$ _{2}$. Our findings demonstrate an ideal platform to implement applications on topological valleytronics.
\end{abstract}

\maketitle
\section{\label{sec:1}Introduction}
The honeycomb lattice materials, such as graphene \cite{graphene_01,graphene_02,graphene_03,graphene_04} and transition metal dichalcogenides~(TMDs) \cite{tmd_01,tmd_02,tmd_03}, host two inequivalent valleys $ K $ and $K^{ \prime }$ in momentum space located at the corners of the hexagonal Brillouin zone. In addition to the spin of electron, valley is another degree of freedom in two-dimensional honeycomb-like materials. The valley degree of freedom can be tuned by electric fields~\cite{electric1,electric2,electric3}, external magnetic fields~\cite{magnetic1,magnetic2,magnetic3}, and circularly polarized light excitation~\cite{optic1,optic2,optic3}. Similar to spintronics, valleytronics is mainly focused on manipulating valley degree of freedom in designing functional nanodevices~\cite{valley1,valley2,valley3,valley4}, such as electron beam splitter, valley filter, and valley valves\cite{application0,application1,application2,application3,application4}.

Quantum anomalous Hall effect (QAHE) is the quantized version of Hall effect without applying external magnetic fields~\cite{QAHE1,QAHE2,QAHE3,QAHE4}, and has been experimentally realized in various material systems~\cite{QAHE-FM1,QAHE-FM2,QAHE-FM3,QAHE-FM4}. The valley polarization, e.g., in TMD materials, is often topologically trivial~\cite{lifting-valley1,lifting-valley2,lifting-valley3,lifting-valley4}, whereas QAHE is usually a valley degenerated state. To implement valley-polarized QAHE, the inversion symmetry and time-reversal symmetry should be simultaneously broken. One rational way to realize valley-polarized QAHE is by inducing the magnetic proximity effect in the $\mathbb{Z}_2$ topological insulator/ferromagnetic insulator heterostructures~\cite{2021_zheng,V-QAHE1,V-QAHE2,V-QAHE3,2022_majeed,2022_majeed_02}. Although several valley-polarized QAHE systems have been theoretically proposed~\cite{2021_zheng,V-QAHE1,V-QAHE2,V-QAHE3,2022_majeed,2022_majeed_02,V-QAHE-pri-1,V-QAHE-pri-2,2022_zheng_high_order,2022_wangrui}, more experimentally realizable candidates are still highly desired to observe the valley-polarized QAHE.

In this article, we propose an ideal platform to realize valley polarization and valley-polarized QAHE based on monolayer jacutingaite family materials. Jacutingaite (Pt$_{2}$HgSe$_{3}$) is a natural mineral with large intrinsic spin-orbit coupling (SOC). The atomic structure of jacutingaite can be treated as a quarter of $ X $ atoms ($ X =$ S, Se, Te) of the 1$T$-phase TMD Pt$X_{2}$ replaced by $ A $ atoms ($ A =$ Cd, Hg, Zn), and the monolayer is crystallized in $ P\overline{3}m1 $ space group~(No.~164) with a buckled honeycomb lattice.
The dynamically stable monolayer Pt$_{2}$HgSe$_{3}$ and its family materials Pt$ _{2}AX_{3} $ ($ A $ = Hg, Cd, Zn; $ X $ = S, Se, Te) are $\mathbb{Z}_{2}$ topological insulators~\cite{Exp_PHS1,Exp_PHS2,PHS1,PHS2,PHS3,Kane-Mele,PHS4,DFT_PHS3}. The monolayer ferromagnetic substrates, i.e. $ MY_{2} $ ($ M = $ Fe, Co, Ni; $ Y = $ Cl, Br, I), CI$ _{3} $, VI$ _{3} $, and MnBi$ _{2} $Te$ _{4} $, are used to induce magnetic exchange interactions~\cite{mx2_all,curie_mx2_ph,mx2_CrI3,curie_mbt,addU_Cr_V,addU_Co,addU_Fe_Ni,addU_Mn,mx2_u=2,mx2_NiI2,mx2_VI3,mx2_crx3_strain,mag1,mag2,mag3,mag4}. These two-dimensional magnets possess the ferromagnetic ground states in freestanding monolayer under the Curie temperatures as summarized in Table~S2 \cite{SM}.
By performing first-principles calculations, we systematically investigate the atomic, electronic, and topological properties of over 100 kinds of Pt$ _{2}AX_{3} $/ferromagnetic substrate van der Waals (vdW) heterostructures.
By examining the lattice mismatch, stacking configuration, and band alignment, 44 kinds of well-matched vdW heterostructures are selected.
Further analyses on band structures and topological properties demonstrate that these systems host valley polarization and exhibit different Berry curvature distributions at $ K/K^{\prime}$ valleys.

In particular, we observe valley-polarized QAHE in Pt$ _{2} $HgS$ _{3} $/NiBr$ _{2} $, Pt$_{2}$HgSe$_{3}$/CoBr$_{2}$, Pt$_{2}$HgSe$_{3}$/NiBr$_{2}$, and Pt$_{2}$ZnS$_{3}$/CoBr$_{2}$, with the Chern number of $ \mathcal{C} = \pm 1$. 
We find a large valley splitting of 134.2~meV in Pt$_{2}$HgSe$_{3}$/NiBr$_{2}$ heterostructure and a considerable topological band gap of 58.8~meV in Pt$_{2}$HgS$_{3}$/NiBr$_{2}$ heterostructure.
Both the monolayer jacutingaite family materials and the two-dimensional ferromagnetic substrates are energetically and dynamically stable \cite{PHS3,mx2_all,curie_mx2_ph,mx2_CrI3,curie_mbt,addU_Cr_V}, greatly facilitating the experimental implementation of valley-polarized QAHE. Moreover, the structural stabilities of these Pt$ _{2}AX_{3} $-based vdW ferromagnetic heterostructures are also verified by molecular dynamic simulations \cite{SM}.

\section{\label{sec:2}Calculation methods}
Our first-principles calculations were performed by using the projected augmented-wave method \cite{PAW} as implemented in the Vienna $ab~initio$ simulation package (VASP) \cite{VASP1,VASP2}. The generalized gradient approximation of the Perdew-Burke-Ernzerhof (PBE) type was used to describe the exchange-correlation interaction \cite{PBE}. The vdW interaction was treated by using DFT-D3 functional \cite{DFT-D3}. All atoms were fully relaxed until the Hellmann-Feynman force on each atom was less than 0.01~eV/\AA. A vacuum buffer layer of 20 {\AA} was used to avoid unnecessary interaction along $ z $ direction between adjacent slabs. The plane-wave energy cutoff was set to be 520~eV. The $\mathrm{\Gamma}$-centered Monkhorst-Pack $ k $-point grid of $\mathrm{11\times 11\times 1}$ was adopted in all our calculations.
To deal with the strong correlation effect of 3$ d $ magnetic elements (i.e., Fe, Co, Ni, Cr, V, and Mn), the GGA+$U$ method \cite{addUmethod} was used with the corresponding on-site repulsion energy $U$ and exchange interaction $ J $ ~\cite{addU_Cr_V,addU_Co,addU_Fe_Ni,addU_Mn,mx2_u=2} as described in Table~S1 in supplemental materials \cite{SM}. Topological properties were calculated by using maximally-localized Wannier functions as implemented in Wannier90 and WannierTools software packages \cite{wannier90,wannier90new,wanniertools}. Spin-resolved band structures are extracted by using \textsc{vaspkit} software \cite{vaspkit} and \textsc{pyprocar} code \cite{pyprocar}. Atomic structure and charge density difference are visualized by using \textsc{vesta} software package \cite{VESTA}.

\section{\label{sec:3}Atomic structures}
\begin{figure}[tbp]
	\centering
	\includegraphics[width=0.5\textwidth]{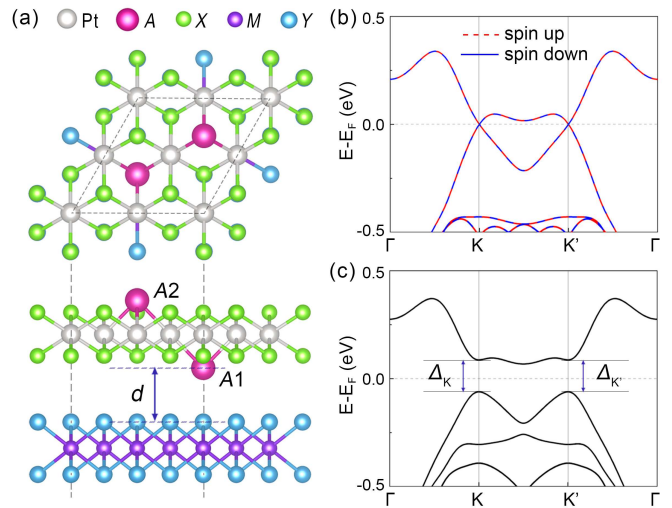}
	\caption{(a) Top and side views of Pt$ _{2}AX_{3} $/$ MY_{2}$ heterostructure where $ 2 \times 2 $ $ MY_{2}$ is adopted to match $ 1 \times 1 $ Pt$ _{2}AX_{3} $. $d$ denotes the optimized vdW gap between the ferromagnetic substrate and jacutingaite family monolayer. $ A $1 and $ A $2 represent different sublattices of $ A $ atoms, where $ A $1 locates closer to the ferromagnetic substrate. (b), (c) Band structures for the pristine jacutingaite Pt$ _{2}$HgSe$_{3} $ monolayer (b) without and (c) with SOC. $\Delta _{K}$ and $\Delta _{K^{\prime}} $ are local band gaps at valley $K$ and $K^\prime$ valleys, respectively.}
	\label{Fig1-atom}
\end{figure}

Monolayer Pt$ _{2}AX_{3}$ ($ A $ = Hg, Cd, Zn; $ X $ = S, Se, Te) is crystallized in $ P\overline{3}m1 $ space group~(No.~164) with a buckled honeycomb structure as shown in Fig.~\ref{Fig1-atom}(a). The $A$ atoms are located at two different sublattices $ A1 $ and $ A2 $.
The jacutingaite family monolayer Pt$ _{2}AX_{3}$ belongs to Kane-Mele type topological insulator \cite{Kane-Mele,PHS3}, with the maximum band gap of 178~meV for Pt$ _{2}$ZnSe$ _{3}$~\cite{PHS3}. The typical band structures of monolayer Pt$_{2}AX_{3}$ are displayed in Figs.~\ref{Fig1-atom}(b) and \ref{Fig1-atom}(c), where the Dirac points can be observed at $ K $ and $K ^{\prime}$ valleys in the absence of SOC [see Fig.~\ref{Fig1-atom}(b)]. When the SOC is considered, as shown in Fig.~\ref{Fig1-atom}(c), a topological nontrivial band gap opens up around the two valleys. The valley degeneracy of monolayer Pt$ _{2}AX_{3}$ is protected by the time-reversal symmetry and inversion symmetry, leading to $ \Delta_{K}  = \Delta_{{K}^{\prime}} $.

\begin{figure*}
	\centering
	\includegraphics[width=1.0\textwidth]{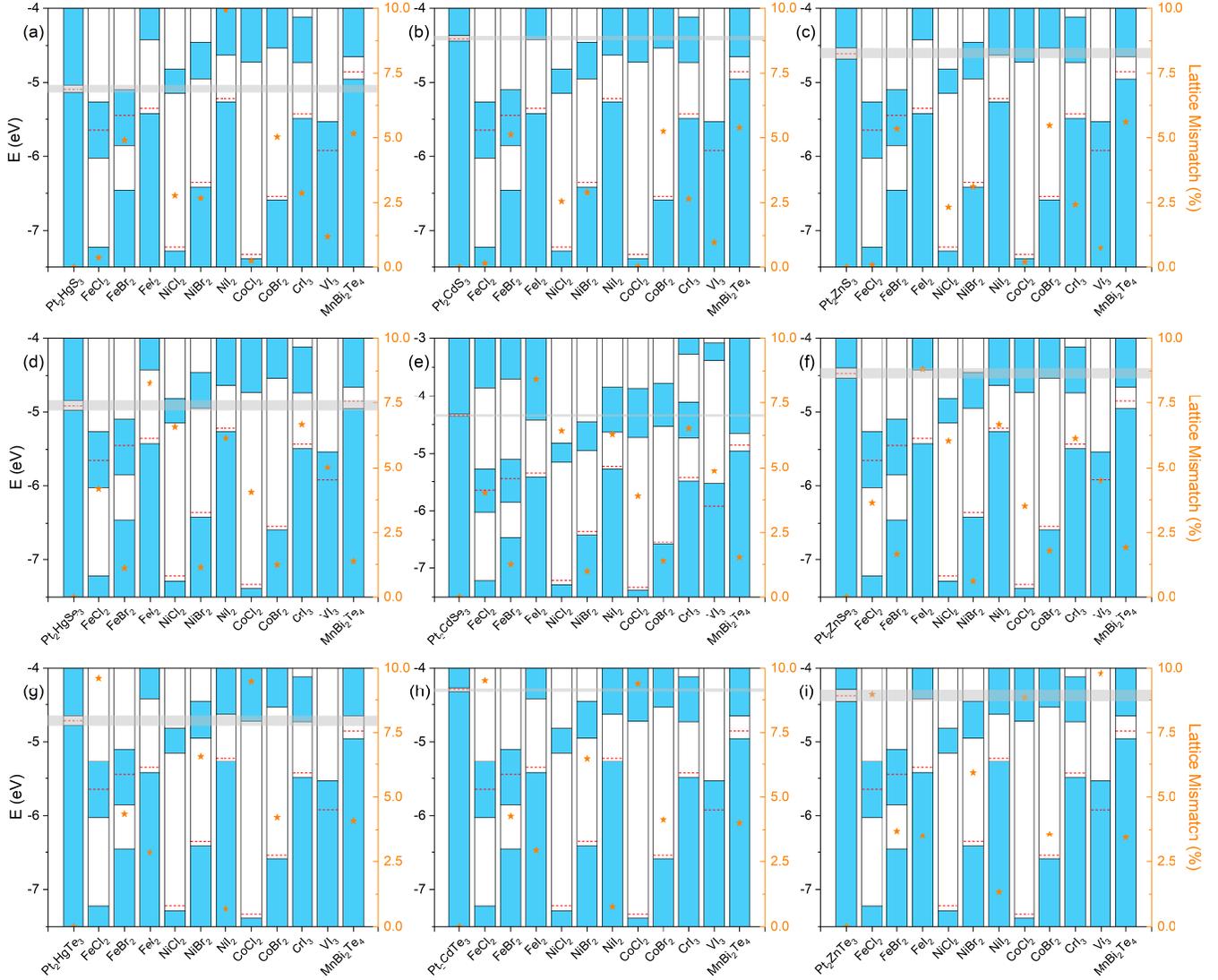}
	\caption{Band alignments and lattice mismatch of jacutingaite family monolayer Pt$ _{2}AX_{3} $/ferromagnetic substrate heterostructures. Gray region stands for the intrinsic band gap of each jacutingaite family monolayer Pt$ _{2}AX_{3} $. The blue (white) area represents the presence (absence) of electronic states. Band edges are relative to the vacuum level by calculating work functions (dashed red line). Orange star (right axis) denotes the lattice mismatch of corresponding vdW heterostructure. (a)-(c), (d)-(f), and (g)-(i) represent the cases with $ X $~= S, Se, Te, respectively. (a)(d)(g), (b)(e)(h), and (c)(f)(i) represent the cases with $ A $~= Hg, Cd, Zn, respectively.}
	\label{Fig2-align}
\end{figure*}

In our calculations, we adopt $ 1 \times 1 $ Pt$ _{2}AX_{3} $ and different ferromagnetic substrates with suitable supercell to match the lattice constant. For example, the vdW heterostructure as shown in Fig.~\ref{Fig1-atom}(a) consists of $ 1 \times 1 $ Pt$ _{2}AX_{3} $ and  $ 2 \times 2 $ $ MY_{2}$. To find the well-matched vdW heterostructures, we carry out the lattice mismatch calculations as plotted in Fig.~\ref{Fig2-align} with orange stars and summarized in Table~S3 in supplemental materials \cite{SM}. Considering the experimental feasibility, we mainly explore the physical properties of 62 kinds of heterostructures with lattice mismatch less than $ 5\% $ as highlighted in Table~S3~\cite{SM}. We then investigate the most stable stacking orders of these vdW heterostructures. For Pt$ _{2}AX_{3} $/$ MY_{2}$, we consider four stacking orders according to the atom site directly below \textit{A}1 atoms: (i)~$ M $, (ii)~$ Y $(top), (iii)~$ Y $(hollow), and (iv)~$ MY $(bridge), respectively. Fig.~\ref{Fig1-atom}(a) displays the type-i stacking configuration and the other three configurations are displayed in Fig.~S3 in supplemental materials~\cite{SM}. After geometrical optimization and total energy computation for different stacking orders, we summarize the most stable stacking configurations for every well-matched vdW heterostructure in Table~S5 in supplemental materials \cite{SM}.

\section{\label{sec:4}Band alignments}
To design functional devices based on heterostructures, band alignment is a crucial strategy that is widely used to analyze band edges of the two different materials forming a heterostructure~\cite{alignment1,alignment2}. According to the different band gap positions, band alignment can be specified into three types: (I) straddling gap, (II) staggered gap, and (III) broken gap. Due to the type-III band alignment leading to the absence of global band gap in a heterostructure \cite{typeIII-nogap-1,typeIII-nogap-2}, we apply the band alignment strategy to screen out the type-I and type-II vdW heterostructures in order to obtain a global band gap.

\begin{table*}
	\centering
	\caption{Transferred charge $ \Delta Q $, vdW gap $ d $, and magnetic moments of 8 kinds of typical Pt$ _{2}AX_{3} $/$ MY_{2}$ vdW heterostructures. $ \Delta Q $ is extracted from the Bader charge analysis to evaluate the transferred electrons from the ferromagnetic substrate to Pt$ _{2}AX_{3} $ layer. $ d $ represents the interlayer distance after full relaxation. $ M_{\text{jac}} $, $ M_{\text{tot}} $, $ M_{A1} $, and $ M_{A2} $ denote the magnetic moment of Pt$ _{2}AX_{3} $ layer, total vdW heterostructure, $ A1 $ atom, and $ A2 $ atom, respectively.}
	\renewcommand\arraystretch{1.1}
	\begin{ruledtabular}
		\begin{tabular}{lcccccccc}
			\multirow{2}{*}{\shortstack{vdW \\ Heterostructures}}	&		\multirow{2}{*}{\shortstack{Pt$_{2}$HgS$_{3}$/ \\ CoBr$_{2}$} } 	&	\multirow{2}{*}{ \shortstack{	Pt$_{2}$HgSe$_{3}$/ \\ CoBr$_{2}$ }}&	\multirow{2}{*}{ \shortstack{	Pt$_{2}$HgTe$_{3}$/ \\ CoBr$_{2}$  }}	&		\multirow{2}{*}{ \shortstack{Pt$_{2}$ZnS$_{3}$/ \\ CoBr$_{2}$ }}	&	\multirow{2}{*}{ \shortstack{	Pt$_{2}$HgS$_{3}$/ \\ NiBr$_{2}$  }}	&	\multirow{2}{*}{ \shortstack{Pt$_{2}$HgSe$_{3}$/ \\ NiBr$_{2}$ }}	&	\multirow{2}{*}{ \shortstack{	Pt$_{2}$CdS$_{3}$/ \\ NiBr$_{2}$ }} & \multirow{2}{*}{ \shortstack{Pt$_{2}$ZnSe$_{3}$/ \\ NiBr$_{2}$ }	}\\
			& &	& & & & & &  \\
			\hline
			$ \Delta Q $~($ e $)	&	-0.17 	&	-0.20 	&	-0.09 	&	-0.32 	&	-0.03 	&	-0.11 	&	-0.18 	&	-0.17 	\\
			$ d $~(\AA)	&	2.51 	&	2.74 	&	2.90 	&	2.70 	&	2.28 	&	2.46 	&	2.18 	&	2.51 	\\
			$ M_{\text{jac}} $~$ (\mu_{\text{B}}) $	&	0.330 	&	0.061 	&	0.068 	&	0.050 	&	0.219 	&	0.204 	&	0.276 	&	0.167 	\\
			$ M_{\text{tot}} $~$ (\mu_{\text{B}}) $	 &	11.525 	&	11.530 	&	11.490 	&	11.503 	&	7.453 	&	7.446 	&	7.392 	&	7.420 	\\
			$ M_{A1}  $~$ (\mu_{\text{B}}) $	&	0.005 	&	-0.009 	&	-0.005 	&	-0.014 	&	-0.016 	&	-0.021 	&	-0.020 	&	0.009 	\\
			$ M_{A2}  $~$ (\mu_{\text{B}}) $	&	0.056 	&	0.016 	&	0.007 	&	0.024 	&	0.067 	&	0.045 	&	0.104 	&	0.027 	\\
		\end{tabular}
	\end{ruledtabular}
	\label{tab-charge}
\end{table*}

Figure~\ref{Fig2-align} displays the band alignments of jacutingaite family monolayer Pt$ _{2}AX_{3} $ with different ferromagnetic substrates, which are obtained by calculating the work function and band structures of each freestanding material. We first take Pt$ _{2}$HgS$_{3} $ as an example to analyze the band alignment. The first column in Fig.~\ref{Fig2-align}(a) represents the band edge of monolayer Pt$ _{2}$HgS$_{3} $, and the Fermi level (dashed red line) with respect to the vacuum level (0~eV) is about $ -5 $~eV lying inside the intrinsic band gap as displayed in gray region. The blue (white) area represents the presence (absence) of electronic states. One can find that the band gap of Pt$ _{2}$HgS$_{3}$ layer lies in the band gap of NiBr$ _{2}$, resulting in a type-I heterostructure of Pt$ _{2}$HgS$_{3} $/NiBr$ _{2}$. Similarly, the Pt$ _{2}$HgS$_{3} $/FeBr$ _{2}$ system follows the type-II band alignment with the staggered gap. However, the Pt$ _{2}$HgS$_{3} $/NiCl$ _{2}$ system demonstrates the type-III nature because of the broken gap alignment, which will be excluded in our following discussions.
By combing the band alignment and lattice mismatch analyses, we can filter out 44 kinds of vdW heterostructures from over 100 combinations as summarized in Table~S5 in supplemental materials \cite{SM}, in which we hopefully obtain global band gaps with Chern insulating phase after these heterostructures are formed.

\section{\label{sec:5}Interfacial Electronic and Magnetic Properties}
When two different materials form heterostructures, the charge transfer process usually exists and further modifies the electronic and magnetic properties of the heterostructure \cite{fop_hu2018,fop_yan2018,fop_wang2021}. To explore the interfacial electronic properties of the Pt$ _{2}AX_{3} $/ferromagnetic insulator vdW heterostructures, we calculate the charge density difference and the planar-averaged electrostatic potential along $ z $ direction.
Here, we take eight typical systems with CoBr$_2$ and NiBr$_2$ substrates to illustrate the charge transfer process as shown in Fig.~\ref{Fig3-charge}, Table~\ref{tab-charge} and Figs.~S10-S11 in supplemental materials \cite{SM}.

For Pt$ _{2}$HgSe$_{3} $/NiBr$_{2}$, as displayed in Fig.~\ref{Fig3-charge}(a), the planar-averaged electrostatic potential alone $ z $ direction in Pt$ _{2}$HgSe$_{3} $ layer is no longer symmetric with respect to the Pt atomic layer, indicating the presence of charge transfer process at the interface. More quantitatively, the Bader charge analysis demonstrates the transferred charge $ \Delta Q $ from NiBr$_{2}$ layer to Pt$ _{2}$HgSe$_{3} $ layer is $ -0.11 e $. The charge density difference, as illustrated in Fig.~\ref{Fig3-charge}(b), also demonstrates that the NiBr$_{2}$ layer accumulates electrons whereas the Pt$ _{2}$HgSe$_{3} $ layer loses electrons after they form a heterostructure.
As a result, a built-in electric field $ \Delta V_{z} $ appears at the interface as displayed in Fig.~\ref{Fig3-charge}(a), which can further enhance the Rashba SOC of the Pt$ _{2}$HgSe$_{3} $/NiBr$_{2}$ system.

\begin{figure}[tbp]
	\centering
	\includegraphics[width=0.5\textwidth]{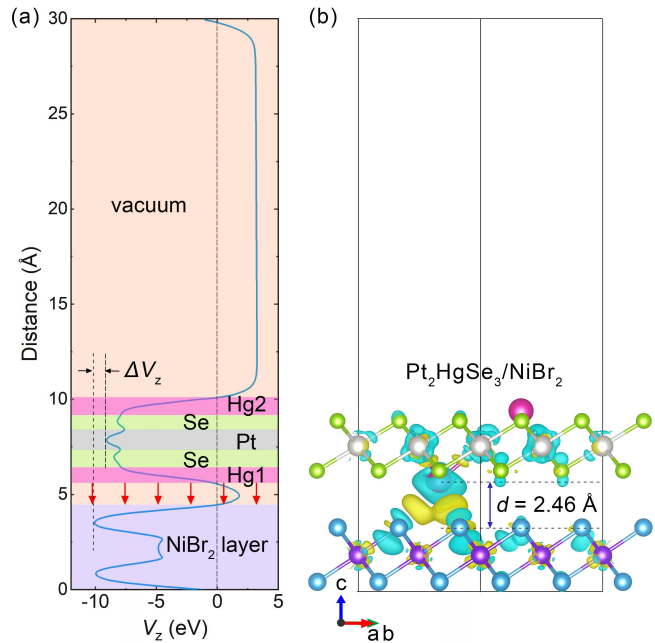}
	\caption{ (a) The planar-averaged electrostatic potential $ V_{z} $ and (b) charge density difference of the Pt$ _{2}$HgSe$_{3} $/NiBr$_{2}$ heterostructure. $ V_{z} $ in Pt$ _{2}$HgSe$_{3} $ layer is asymmetric with respect to the Pt atomic layer, indicating the charge transfer at the interface. Highlighted regions in (a) correspond to associated horizontal atomic layers in (b). Cyan and yellow contours represent charge depletion and accumulation, respectively.}
	\label{Fig3-charge}
\end{figure}

\begin{figure*}
	\centering
	\includegraphics[width=1\textwidth]{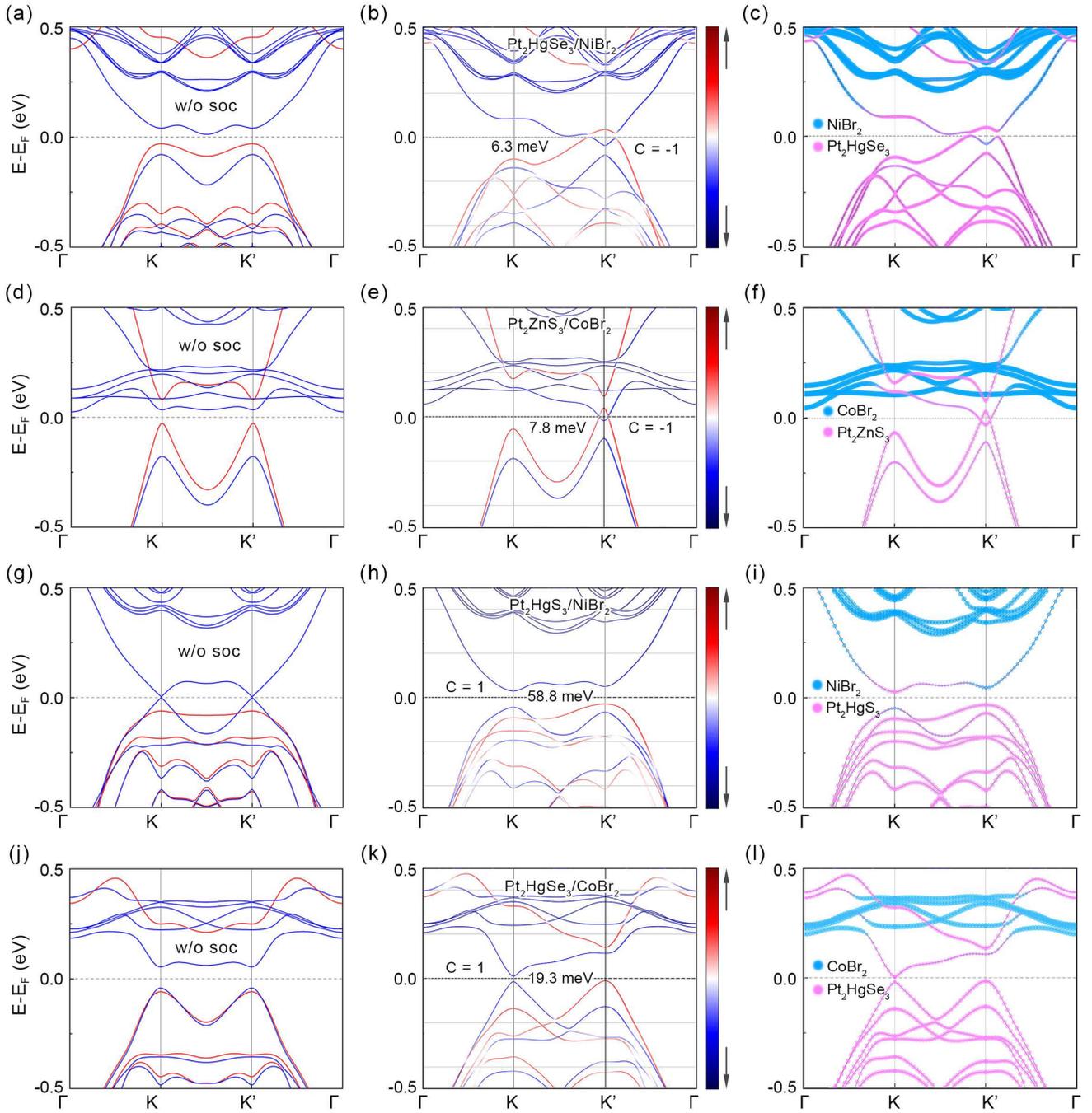}
	\caption{(a)-(b) Spin-resolved band structures of the Pt$ _{2}$HgSe$_{3} $/NiBr$_{2}$ heterostructure (a) without and (b) with SOC, respectively. (d)-(e) Spin-resolved band structures of the Pt$_{2}$ZnS$_{3}$/CoBr$_{2}$ heterostructure (d) without and (e) with SOC, respectively.  (g)-(h) Spin-resolved band structures of the Pt$ _{2}$HgS$_{3} $/NiBr$_{2}$ heterostructure (g) without and (h) with SOC, respectively.  (j)-(k) Spin-resolved band structures of the Pt$ _{2}$HgSe$_{3} $/CoBr$_{2}$ heterostructure (j) without and (k) with SOC, respectively. The red (blue) color denotes the spin-up (spin-down) state. The Chern number $\mathcal{C}$  and the global band gap for every system is demonstrated in (b), (e), (h), and (k), respectively. (c), (f), (i), (l) Layer-resolved band structures of (c) Pt$ _{2}$HgSe$_{3} $/NiBr$_{2}$, (f) Pt$_{2}$ZnS$_{3}$/CoBr$_{2}$, (i) Pt$ _{2}$HgS$_{3} $/NiBr$_{2}$, and (l) Pt$_{2}$HgSe$_{3}$/CoBr$_{2}$. The pink bubble represents the element projection of jacutingaite family layer whereas the light blue bubble represents the element projection of ferromagnetic substrate layer.}
	\label{Fig4-bands}
\end{figure*}

Table~\ref{tab-charge} displays the transferred charge $ \Delta Q $, vdW gap $ d $, and magnetic moments of the eight typical Pt$ _{2}AX_{3} $/$ MY_{2}$ vdW heterostructures. One can observe that all the interlayer vdW gaps $d$ become smaller than the initial state after full structural relaxation. Due to the smaller atomic radius of S atom compared to Se atom, Pt$ _{2}$HgS$_{3} $ tends to move closer to the substrate than Pt$ _{2}$HgSe$_{3} $ when constructing a heterostructure with the same substrate. Moreover, $ d $ in Pt$ _{2}AX_{3} $/NiBr$_2$ is generally smaller than that in Pt$ _{2}AX_{3} $/CoBr$_2$. 
The shorter $d$ in Pt$ _{2}AX_{3} $/NiBr$_2$ leads to a stronger magnetic proximity effect and large magnetization $ M_{\text{jac}} $ in jacutingaite layer. The magnetic proximity effect also induces different magnetic moments $ M_{A1} $ and $ M_{A2} $ on $A1$ and $A2$ because of the different distances to the substrate. The presence of a the ferromagnetic substrate breaks both inversion and time-reversal symmetries and leads to inequivalent exchange fields on $A1$ and $A2$ atoms, which lifts the spin degeneracy and valley-dependent band gaps. Nonzero Berry curvature and anomalous Hall responses can also appear \cite{2021_zheng}.
For example, in Pt$ _{2}$HgSe$_{3}$/NiBr$_{2}$, the magnetic moment of Hg2 atom is 0.045 $\mu_{B}$ and is parallel to that of NiBr$_{2}$, whereas an antiparallel magnetic moment of -0.021 $\mu_{B}$ is induced in Hg1 atom. Consequently, the valley degeneracy of Pt$ _{2}AX_{3} $ can be lifted by the proximity effect induced broken time-reversal symmetry and Rashba SOC. 
Hereinbelow, we explore the electronic properties of these Pt$ _{2}AX_{3} $/ferromagnetic substrate vdW heterostructures.

\section{\label{sec:6}Electronic and Topological Properties}
After calculating band structures and topological properties of the 44 kinds of well-matched Pt$ _{2}AX_{3} $/ferromagnetic substrate vdW heterostructures, we find four systems are topologically nontrivial, whereas the remaining systems are topologically trivial (see Figs.~S12-S22 in supplemental materials \cite{SM} for band structures of all 44 systems).
Figure~\ref{Fig4-bands} shows the spin-resolved and layer-resolved band structures of the four topologically nontrivial systems, i.e., Pt$ _{2}$HgSe$_{3} $/NiBr$_{2}$, Pt$_{2}$ZnS$_{3}$/CoBr$_{2}$, Pt$ _{2}$HgS$_{3} $/NiBr$_{2}$, and Pt$_{2}$HgSe$_{3}$/CoBr$_{2}$.

We start from Pt$ _{2}$HgSe$_{3} $/NiBr$_{2}$ as an example [see Figs.~\ref{Fig4-bands}(a)-\ref{Fig4-bands}(c)].
In the absence of SOC, as shown in Fig.~\ref{Fig4-bands}(a), we can observe that the magnetic proximity effect of ferromagnetic substrates induces sizable Zeeman splitting but keeps the valley degeneracy. Further calculations on the band structures with and without magnetization also demonstrate that the band structures of the Pt$ _{2} $AX$ _{3} $ near the Fermi level are modified by the magnetic exchange interaction \cite{SM}. When SOC is considered, as displayed in Fig.~\ref{Fig4-bands}(b), a band inversion appears around $K^\prime$ point with a band gap of 6.3~meV induced by Rashba SOC, whereas the bands close to the Fermi level around $K$ point move far away from each other in a repulsive manner, forming a sizable local band gap of $\Delta_{K}$ of 182.8~meV. The difference around $K/K^\prime$ valleys results in a large valley splitting of 134.2 meV.
We can also observe from the layer-resolved band structures as shown in Fig.~\ref{Fig4-bands}(c) that the electronic states near the Fermi level are mainly dominated by the topological Pt$ _{2}$HgSe$_{3}$ layer, indicating the electronic structures of the Pt$ _{2}$HgSe$_{3}$ monolayer are modified after coupled to the NiBr$_{2}$ substrate.

The above band analysis can be applied to the remaining three nontrivial systems. For Pt$ _{2}$ZnS$_{3} $/CoBr$_{2}$ system, as shown in Figs.~\ref{Fig4-bands}(d)-\ref{Fig4-bands}(f), the band structures are similar to that of Pt$ _{2}$HgSe$_{3} $/NiBr$_{2}$ with a band inversion and a global band gap of 7.8~meV around $K^\prime$ point induced by Rashba SOC.
For Pt$ _{2}$HgS$_{3} $/NiBr$_{2}$ and Pt$_{2}$HgSe$_{3}$/CoBr$_{2}$ systems, as illustrated in Figs.~\ref{Fig4-bands}(g)-\ref{Fig4-bands}(l), the bands are different from that of Pt$ _{2}$HgSe$_{3} $/NiBr$_{2}$, where the local gap at $K$ valley is smaller than that at $K^\prime$ valley. Table~\ref{tab-nontrivial} summarizes the extracted valley splitting $ \Delta $, global band gaps $ E_{\text{g}} $, local band gaps $ \Delta _{K}  $/$ \Delta _{K^{\prime}} $ at valleys $K/K^\prime$, and Chern numbers $\mathcal{C}$ for above four vdW heterostructures. One can see that the valley polarization generally exists in these systems (also see Table~S5 in supplemental materials \cite{SM} for the remaining 40 kinds of heterostructures).
Among the 44 systems, the Pt$_{2}$HgSe$_{3}$/NiBr$_{2}$ exhibits the maximum valley splitting of 134.2~meV, which is an order of magnitude larger than experimental observations in TMD/ferromagnetic substrate systems \cite{TMD_valley_WSe2,TMD_valley_WS2}, providing an ideal platform to realize topological valleytronics.

\begin{table}
	\caption{Valley splitting $ \Delta $, global band gap $ E_{\text{g}} $, local band gaps $ \Delta _{K}  $/$ \Delta _{K^{\prime}} $ at $K/K^\prime$ valleys, and Chern number $\mathcal{C}$ for nontrivial jacutingaite family monolayer/ferromagnetic substrate vdW heterostructures. Valley splitting is evaluated via the difference between valence band maximum at valley $K/K^\prime$, i.e. $\Delta = E_{\text{V}}({K^{\prime}}) - E_{\text{V}}({K})$. }
	\renewcommand\arraystretch{1.3}
	\begin{ruledtabular}
		\begin{tabular}{lrrrrr}
			\multirow{2}{*}{\centering \shortstack{Nontrivial vdW \\ Heterostructrues}} &
			\multirow{2}{*}{\centering \shortstack{$ \Delta $\\(meV)}} &
			\multirow{2}{*}{\centering \shortstack{$ E_{\text{g}} $\\(meV)}}   &
			\multirow{2}{*}{\centering \shortstack{$ \Delta _{K}  $\\(meV)}} &
			\multirow{2}{*}{\centering \shortstack{$ \Delta _{K^{\prime}} $\\(meV)}}   &
			\multirow{2}{*}{$\mathcal{C}$ }  \\
			& & & & &  \\
			\hline
			Pt$_{2}$HgSe$_{3}$/NiBr$_{2}$ 	&  134.2 &  	6.3 &	182.8 & 73.5 & 	-1        \\
			Pt$_{2}$ZnS$_{3}$/CoBr$_{2}$ 	&  36.3  &  7.8   &  185.9 &59.9 	& -1      \\
			Pt$_{2}$HgS$_{3}$/NiBr$_{2}$ 	&   15.4 &  58.8 &	74.2 	&78.8 	& 1   \\
			Pt$_{2}$HgSe$_{3}$/CoBr$_{2}$	&  3.5 	 &   19.3 	  &  22.8 	&124.0 & 	1     \\
		\end{tabular}
	\end{ruledtabular}
	\label{tab-nontrivial}
\end{table}

\begin{figure}
	\centering
	\includegraphics[width=0.5\textwidth]{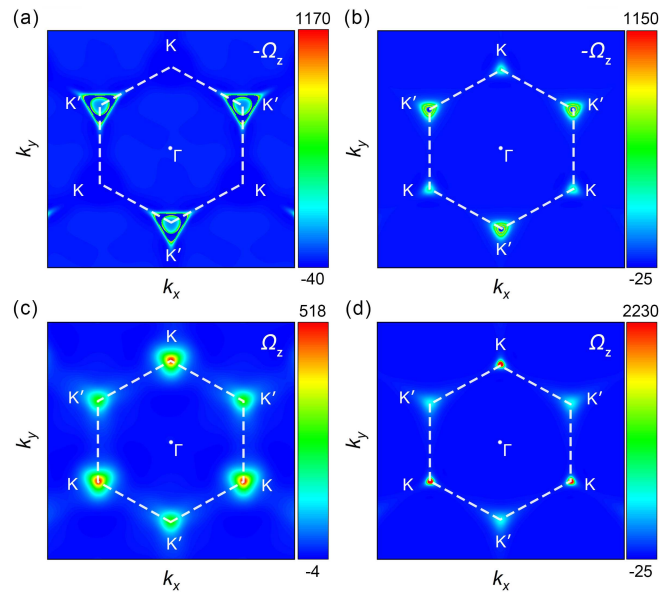}
	\caption{Berry curvature distributions in the first Brillouin zone of four nontrivial vdW heterostructures (a) Pt$ _{2} $HgSe$ _{3} $/NiBr$ _{2} $, (b) Pt$ _{2} $ZnS$ _{3} $/CoBr$ _{2} $, (c) Pt$ _{2} $HgS$ _{3} $/NiBr$ _{2} $, and (d) Pt$ _{2} $HgSe$ _{3} $/CoBr$ _{2} $, respectively. Berry curvature is centered around the ${K}/{K}^{\prime}$ valleys and is absent elsewhere but dissimilarly distributes in the two inequivalent valleys $K/K^{\prime}$. For clarity, (a) and (b) plot $ - \Omega_{z}$ whereas (c) and (d) plot $\Omega_{z}$.}
	\label{Fig5-Berry}	
\end{figure}

\begin{figure*}
	\centering
	\includegraphics[width=0.9\textwidth]{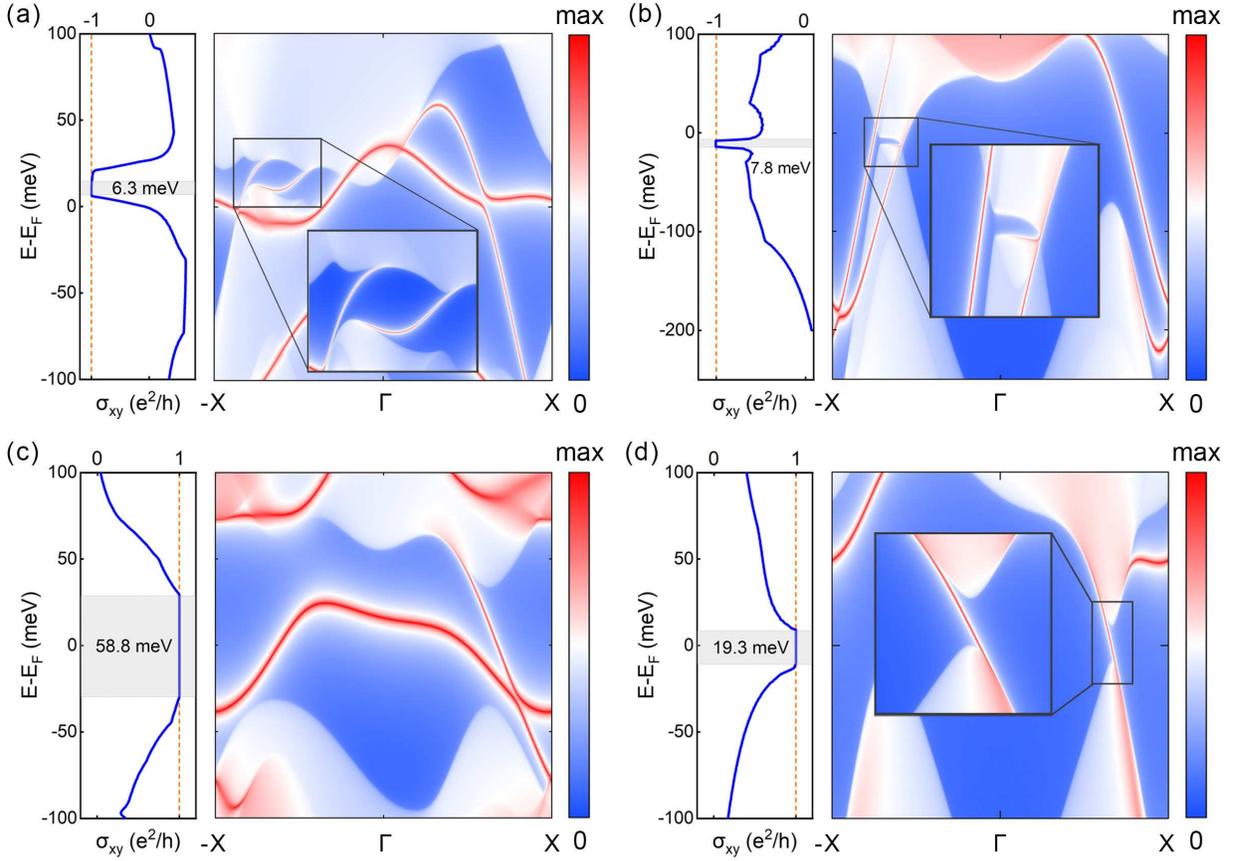}
	\caption{Anomalous Hall conductance $ \sigma_{xy} $ and local density of states (LDOS) of four nontrivial vdW heterostructures (a) Pt$ _{2} $HgSe$ _{3} $/NiBr$ _{2} $, (b) Pt$ _{2} $ZnS$ _{3} $/CoBr$ _{2} $, (c) Pt$ _{2} $HgS$ _{3} $/NiBr$ _{2} $, and (d) Pt$ _{2} $HgSe$ _{3} $/CoBr$ _{2} $. Insets correspond to enlarged views of edge states. In all LDOS plots, there are chiral edge states connecting conduction bands and valence bands.}
	\label{Fig6-LDOS}
\end{figure*}

To evaluate the topology of the above band structures, we compute the Berry curvature of all occupied bands in each system by using maximally localized Wannier functions \cite{wannier90,wannier90new}. We carry out orbital projections for every insulating system to guide the Wannier functions. For example, by analyzing the orbital projection on each kind of elements Pt$ _{2} $HgSe$ _{3} $/NiBr$ _{2} $, we find that Pt-$ d $, Hg-$ s $, Se-$ p $, Ni-$ d $, and Br-$ p $ orbitals are distributed around Fermi level (see Figs.~S1-S2 in supplemental materials \cite{SM}).
Figure~\ref{Fig5-Berry} displays the Berry curvature distributions of above four systems in the first Brillouin zone. One can see that the Berry curvature is centered around the ${K}/{K}^{\prime}$ valleys but is absent elsewhere.
The value of the Berry curvature around $K/K^\prime$ valleys depends on the specific material combinations, i.e., the Berry curvature only appears in $K^\prime$ valley for Pt$ _{2} $HgSe$ _{3} $/NiBr$ _{2} $ as displayed in Fig.~\ref{Fig5-Berry}(a), while it exists in both $K$ and $K^\prime$ valleys with different magnitudes for other three systems as shown in  Figs.~\ref{Fig5-Berry}(b)-\ref{Fig5-Berry}(d).
By integrating the Berry curvature all over the first Brillouin zone, we can obtain the Chern numbers as listed in Table~\ref{tab-nontrivial}. The nonzero Chern number indicates the nontrivial band topology in the four systems, i.e., the formation of QAHE.
Combining the valley polarizations and different Berry curvature distributions at the two inequivalent valleys, we can find that these four systems can implement valley-polarized QAHE. More strikingly, the global band gap of 58.8~meV in Pt$_{2}$HgS$_{3}$/NiBr$_{2}$ is over three times larger than that in Pt$_{2}$HgSe$_{3}$/CrI$_3$ heterostructure (17.2~meV) as predicted in our previous report~\cite{2021_zheng}, implying the potential for high-temperature QAHE \cite{fop_deng2018,high-temp-1,high-temp-2,high-temp-3,high-temp-4,high-temp-5}. Surprisingly, the magnetocrystalline anisotropy energy (MAE) calculation shows that the MAE of Pt$ _{2} $ZnS$ _{3} $/CoBr$ _{2} $ heterostructure is 1.795~meV/Co, suggesting the magnetic easy axis of CoBr$ _{2} $ can be tuned by Pt$ _{2} $ZnS$ _{3} $ to out-of-plane direction \cite{SM}, which is beneficial to the experimental realization of valley-polarized QAHE.

The nontrivial topology of the four systems can also be confirmed by the anomalous Hall conductance $\sigma_{xy}$ and the local density of states (LDOS) calculations, as displayed in Fig.~\ref{Fig6-LDOS}.
The anomalous Hall conductance in Figs.~\ref{Fig6-LDOS}(a) and \ref{Fig6-LDOS}(b) demonstrates a quantized Hall plateau of $\sigma_{xy}=-e^2/h$  in Pt$ _{2} $HgSe$ _{3} $/NiBr$ _{2} $ and Pt$ _{2} $ZnS$ _{3} $/CoBr$ _{2} $ systems when the Fermi energy lies inside the global band gap, which is a strong evidence for QAHE.
Similarly, in Figs.~\ref{Fig6-LDOS}(c) and \ref{Fig6-LDOS}(d), we can also observe that the anomalous Hall conductance is quantized as $\sigma_{xy}=e^2/h$ in Pt$ _{2} $HgS$ _{3} $/NiBr$ _{2} $ and Pt$ _{2} $HgSe$ _{3} $/CoBr$ _{2} $ systems.
Such a Chern insulating topological phase is also confirmed by the LDOS of these nontrivial systems since there are chiral edge states in the global band gaps connecting conduction bands and valence bands.

\section{\label{sec:7}Summary and conclusion}

In this work, we proposed an excellent platform to implement valley polarization and valley-polarized QAHE based on the vdW heterostructure constructed by monolayer jacutingaite family materials Pt$ _{2}AX_{3} $ and two-dimensional ferromagnetic substrates. We systematically investigated atomic structures, band alignments, interfacial characteristics, electronic and topological properties of over 100 kinds of Pt$ _{2}AX_{3} $/ferromagnetic substrate heterostructures. By using the band alignment strategy, we filtered out 44 kinds of well-matched systems with potential for global band gaps. We evaluated the interfacial characteristics by analyzing the charge density difference, Bader charge transferring, planar-averaged electrostatic potential, and magnetic properties.
The interlayer charge transfer leads to a built-in electric field and the asymmetric atomic structure results in a broken inversion symmetry. Furthermore, the reducing interlayer vdW gap leads to a strong magnetic proximity, which induces large magnetization in Kane-Mele type topological insulator Pt$ _{2}AX_{3} $.
Consequently, the valley degeneracy is lifted by the breaking of time-reversal symmetry and inversion symmetry. We found the general valley polarization in the 44 kinds of well-matched systems and observed a sizable valley splitting of 134.2~meV in Pt$ _{2} $HgSe$ _{3} $/NiBr$ _{2} $ system.

Particularly, we identified four Chern insulators in Pt$ _{2} $HgS$ _{3} $/NiBr$ _{2} $, Pt$_{2}$HgSe$_{3}$/CoBr$_{2}$, Pt$_{2}$HgSe$_{3}$/NiBr$_{2}$, and Pt$_{2}$ZnS$_{3}$/CoBr$_{2}$ due to their nontrivial band topology.
In addition to the valley polarization, we also observed different Berry curvature distributions in the two inequivalent valleys $K/K^{\prime}$ in these four systems, indicating the presence of valley-polarized QAHE.
Eventually, we obtained valley-polarized QAHE with total Chern numbers $ \mathcal{C} = \pm 1$ in the vdW heterostructures based on monolayer jacutingaite family materials. Most surprisingly, we observed a sizable nontrivial global band gap of 58.8~meV in the Pt$ _{2} $HgS$ _{3} $/NiBr$ _{2} $ system with potential implementation for the high-temperature QAHE. 

In conclusion, valley polarization universally exists in the ferromagnetic vdW heterostructures based on monolayer jacutingaite family materials and several valley-polarized QAHE vdW heterostructures can be achieved when the valley splitting is strong enough to lead to a band inversion. Our work proposes an excellent platform based on monolayer jacutingaite family materials Pt$ _{2}AX_{3} $ to implement topological valleytronics and may stimulate more applications on the two-dimensional valley-polarized topological systems.

\begin{acknowledgments}
We are grateful to Prof.~Yang Gao for helpful advice and discussions. This work was financially supported by the National Natural Science Foundation of China (No. 11974327 and No. 12004369), Fundamental Research Funds for the Central Universities (WK3510000010, WK2030020032), Anhui Initiative in Quantum Information Technologies (Grant No.AHY170000). We also thank for the high-performance supercomputing services provided by AM-HPC and the Supercomputing Center of University of Science and Technology of China.
\end{acknowledgments}

\end{document}